\documentclass[aps,12pt,final,notitlepage,oneside,onecolumn,
nobibnotes,nofootinbib,superscriptaddress,noshowpacs,centertags,amsmath,amssymb]{revtex4}
\usepackage{epsfig}
\usepackage{bm}
\usepackage[usenames]{color}

\begin{document}

\title{
Disentangling between $Z'$ and $Z^*$ with first LHC data}

\author{M. V. Chizhov}\thanks{On leave of absence from the
Centre for Space Research and Technologies, Faculty of Physics,
University of Sofia, 1164 Sofia, Bulgaria} \affiliation{Departamento
de F\'isica, Universidade Federal da Para\'iba, Caixa Postal 5008,
Jo\~ao Pessoa, PB 58051-970, Brazil.}


\begin{abstract}
    The resonance production of new chiral spin-1 excited bosons, $Z^*$,
    and their detection through the Drell--Yan process in
    the first physical runs at the CERN LHC are considered.
    The new neutral chiral bosons can be observed as
    a Breit--Wigner resonance peaks
    in the invariant dilepton mass distribution in the same way as
    the well-known hypothetical gauge bosons, $Z'$.
    However, unique new signatures of the chiral bosons exist.
    These signatures could be very important for the interpretation of
    the first LHC data.
    First, there is no Jacobian peak
    in the lepton transverse momentum distribution
    at the kinematical endpoint of the new resonance.
    Second, the lepton angular distribution
    in the Collins--Soper frame for the high on-peak
    invariant masses of the lepton pairs
    has a peculiar ``swallowtail'' shape.
\end{abstract}


\maketitle

\section{Introduction}

Although the LHC has been designed to solve the Higgs puzzle, it is
obvious that the Higgs sector of the electroweak theory cannot be
investigated at the first stage of the LHC running. While the
discovery of the resonantly produced new neutral heavy bosons,
having nonzero decay widths into charged leptons, could be followed
immediately after detectors calibration at the hadron colliders. A
presence of  partons with a broad range of different momenta allows
to flush the whole energetically accessible region, roughly, up to
$\sqrt{s}/6$, where $\sqrt{s}$ is the center-of-mass collider
energy. Even for this year LHC runs, which will be probably at a mere
$\sqrt{s}=10$~TeV, it allows to step above the Tevatron reach.

Two main circumstances designate the Drell--Yan process
$q\bar{q}\to\ell^+\ell^-$ not only as a gold discovery channel, but
also as a precise tool to unveil the resonance properties. First of
all it provides a very low and theoretically well-understood
background at high invariant leptonic masses. The second feature is
related to the  very clean signature of the process and almost
totally reconstructible kinematics\footnote{Up to small individual
transverse momenta of the quarks.}, which allows to investigate the
details of the production and the decay of the new bosons.

In this paper we will consider the case of the resonance production
of new spin-1 heavy bosons and their detection in the Drell--Yan
process using the first CERN LHC data.\footnote{The cases of
spinless and spin-2 bosons can be also incorporated and will be
considered elsewhere.} In the case that such bosons will be observed
as resonance peaks above the $Z$ boson tail in the invariant
dilepton mass distribution, we suggest to investigate in addition
three more experimentally accessible distributions already on the
early stage of the LHC data-taking. These are the differential
distributions as functions of a transverse momentum of the lepton,
its pseudorapidity and the Collins--Soper angle~\cite{CS}. All these
distributions are related to the spin properties of the new boson
and should play crucial role in the analysis of their interactions.

\section{Neutral spin-1 bosons}

New heavy neutral {\color{blue}{\em gauge\/} bosons} are predicted
by many extensions of the Standard Model (SM). They are associated
with additional $U(1)'$ gauge symmetries and are generically called
$Z'$. The {\em gauge\/} interactions of these bosons with matter
lead to a specific angular distribution of the outgoing lepton in
the dilepton center-of-mass reference frame with respect to the
incident parton
\begin{equation}\label{sV}{\color{blue}
    \frac{{\rm d}\sigma_{Z'}}{{\rm d}\cos\theta^*}\propto 1+
    {\rm ASYM}\cdot\cos\theta^*+\cos^2\theta^*},
\end{equation}
which at present is interpreted as a canonical signature for the
intermediate bosons with spin~1. The coefficient ASYM defines the
backward-forward asymmetry, depending on $P$-parity of $Z'$
couplings to fermions. Experimental uncertainties in the sign of
$\cos\theta^*$ in symmetric $pp$ collisions and in the transverse
momenta of the annihilating partons dilute the apparent value of
ASYM and it can be neglected in the ``first data'' analysis.

In addition to the gauge bosons, another type of spin-1 bosons may
exist, which have only a Pauli-like anomalous coupling to fermions
instead of the gauge one. In contrast to the gauge couplings, where
either only left-handed or right-handed fermions participate in the
interactions, the anomalous couplings mix both left-handed and
right-handed fermions and lead to different angular distribution
than Eq.~(\ref{sV}). Therefore, these bosons carry a nonzero {\em
chiral\/} charge, like the Higgs particles.

A corresponding extension of the SM with the new type of spin-1
particles, {\color{red}{\em chiral\/} bosons}, has been proposed in
\cite{MPLA}. According to the symmetry of the SM the new chiral
bosons have been introduced in doublets, just as the Higgs
particles.
In order to compensate the contributions of the new couplings of the
chiral bosons into the chiral anomaly, the doubling of the doublets
both for the chiral and Higgs particles is needed. Besides this, in
order to prevent also the presence of flavor-changing neutral
currents, {\em up\/} and {\em down\/} type fermions should couple to
different doublets of the Higgs and chiral bosons.

Among all introduced new states of the chiral bosons there exist two
heavy neutral $CP$-even and $CP$-odd bosons, which couple to {\em
down\/} type fermions and, therefore, have nonzero decay widths into
charged leptons. These bosons have nearly degenerate masses,
predicted to be slightly above 1~TeV \cite{0609141}. However, it is
impossible to distinguish them in the case of the light final
fermions at the hadron colliders without a spin correlation
analysis. Therefore, in the following we will consider only one type
of these bosons, namely the $CP$-even one.

It has been noted that for the resonance production of the {\em
chiral\/} bosons the original lagrangian in \cite{0609141} can be
substituted by a more simple one for the {\em excited\/} bosons $Z^*$
\begin{equation}\label{Z*ed}{\color{red}
    {\cal L}_{\rm excited}=\frac{g}{2\sqrt{2}M}
    \left(\bar{\ell}\sigma^{\mu\nu}\ell+\bar{d}\sigma^{\mu\nu}d\right)
    \left(\partial_\mu Z^*_\nu-\partial_\nu Z^*_\mu\right)},
\end{equation}
which has been investigated in \cite{CBB}. Here $M$ is the boson
mass and $g$ is the coupling constant of the $SU(2)_W$ weak gauge
group. The bosons, coupled to the tensor quark currents, are some
types of ``excited'' states as far as the only orbital angular
momentum with $L=1$ contributes to the total angular moment, while
the total spin of the system is zero. This property manifests itself
in their derivative couplings to fermions and a different chiral
structure of the interactions in contrast to the gauge ones.

Let us assume for definiteness that the mass of the new $Z^*$ bosons
is equal to 1~TeV,
which is above the Tevatron reach, but they could be reliably
discovered at the LHC even with very low integrated luminosity, less
than 100~pb$^{-1}$. For comparison we will consider topologically
analogous but minimal interactions of the gauge $Z'$ boson
\begin{equation}\label{Z'ed}{\color{blue}
    {\cal L}_{\rm gauge}=\frac{g}{2}
    \left(\bar{\ell}\gamma^{\mu}\ell+\bar{d}\gamma^{\mu}d\right)
    Z'_\mu}
\end{equation}
with the same mass $M$. The coupling constants are chosen in such a
way that all fermionic decay widths in the Born approximation of the
both bosons are identical. It means that their total production
cross sections at the hadron colliders are nearly equal up to
next-to-leading order corrections. Their leptonic decay width
\begin{equation}\label{Gl}
    \Gamma_\ell=\frac{g^2}{48\pi}M\approx 2.8~{\rm GeV}.
\end{equation}
is sufficiently narrow so that they can be identified as resonances
at the hadron colliders in the Drell--Yan process.

\section{Numerical simulations}

\begin{figure}[htb]
\epsfig{file=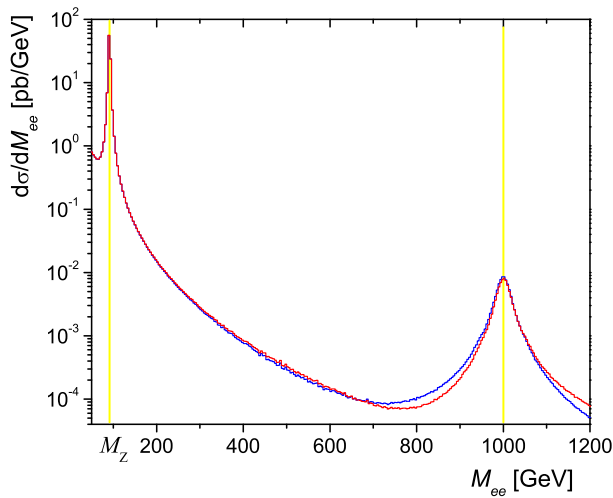,width=9.2cm} \caption{\label{fig:1} The
invariant dilepton mass distributions for the gauge $Z'$ boson
(blue) and the chiral excited $Z^*$ boson (red) with the Drell--Yan
SM background at the CERN LHC.}
\end{figure}

Up to now, the excess in the Drell--Yan process with high-energy
invariant mass of the lepton pairs remains the clearest indication
of the heavy boson production at the hadron colliders. In the
following for the numeric calculations of various distributions we
will use the CalcHEP package~\cite{CalcHEP} with a CTEQ6M choice for
the proton parton distribution set at $\sqrt{s}=10$~TeV. For both
final leptons we impose angular cuts relevant to the LHC detectors
on the pseudorapidity range $|\eta_\ell|<2.5$ and, in addition,
the transverse momentum cuts $p_T > 20$~GeV. The resonance peaks of
new boson production are shown in Fig.~\ref{fig:1}.
It can be concluded that their cross sections at the peak are above two
order of magnitude higher than the corresponding Drell--Yan
background.

\begin{figure}[htb]
\epsfig{file=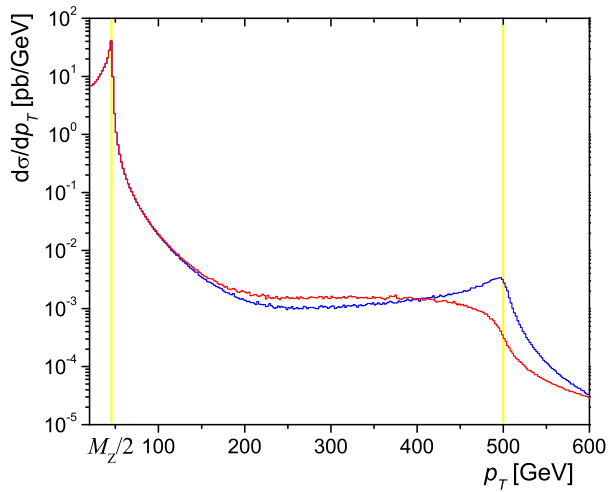,width=9.2cm} \caption{\label{fig:2} The
differential cross sections for the gauge $Z'$ boson (blue) and the
chiral excited $Z^*$ boson (red) with the Drell--Yan SM background
as functions of the lepton transverse momentum at the CERN LHC.}
\end{figure}

The peaks in the invariant mass distributions originate from the
Breit--Wigner propagator form, which is the same both for the gauge
and the chiral bosons in the Born approximation. However, the common
wisdom, that a peak in the invariant mass distribution of the two
final particles must correspond to the Jacobian peaks in their
transverse momentum distributions, is not valid for the chiral
bosons due to the following fact. The main feature of the
interactions (\ref{Z*ed}) consists in different angular distribution
of final fermions~\cite{two}
\begin{equation}\label{sT}{\color{red}
    \frac{{\rm d}\sigma_{Z^*}}{{\rm d}
    \cos\theta^*}\propto \cos^2\theta^*}
\end{equation}
in comparison with the distribution (\ref{sV}) for the gauge
interactions. It leads to a stepwise lepton transverse momentum
distribution, rather than to the Jacobian peak at the kinematical
endpoint $M/2$ for the gauge bosons (Fig.~\ref{fig:2}).
Therefore, already the lepton transverse momentum distribution
demonstrates a difference between the gauge and the chiral bosons.
In order to make more definite conclusions, let us investigate other
distributions selecting only ``on-peak'' events with the invariant
dilepton masses in the range 800~GeV~$<M_{\ell\ell}<$~1200~GeV.

\begin{figure}[th]
\epsfig{file=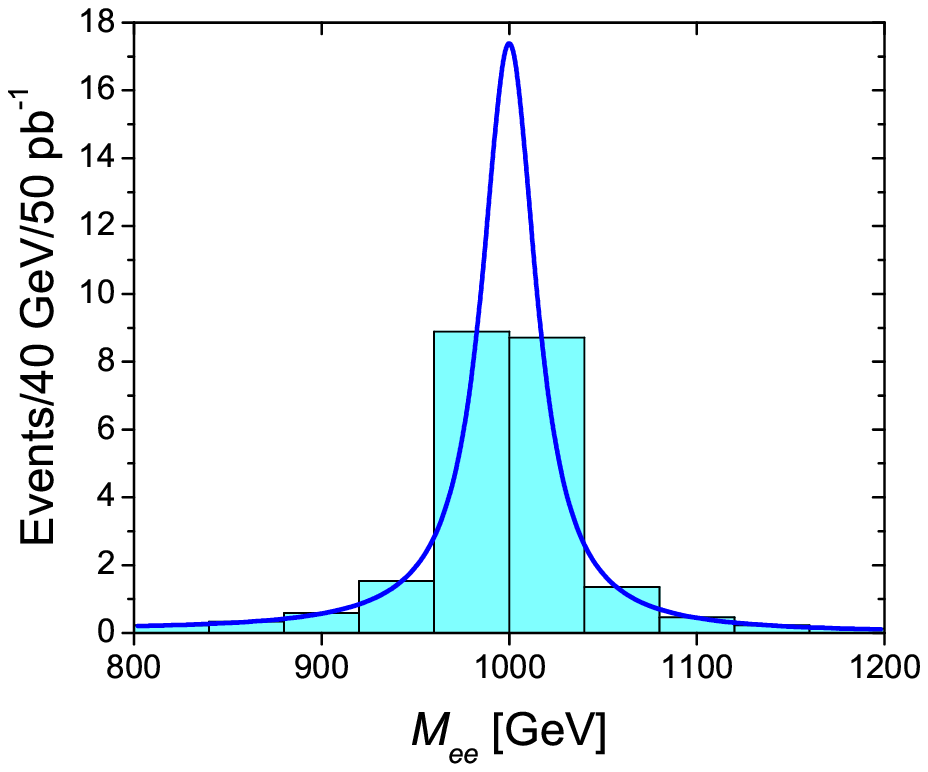,width=8cm}\epsfig{file=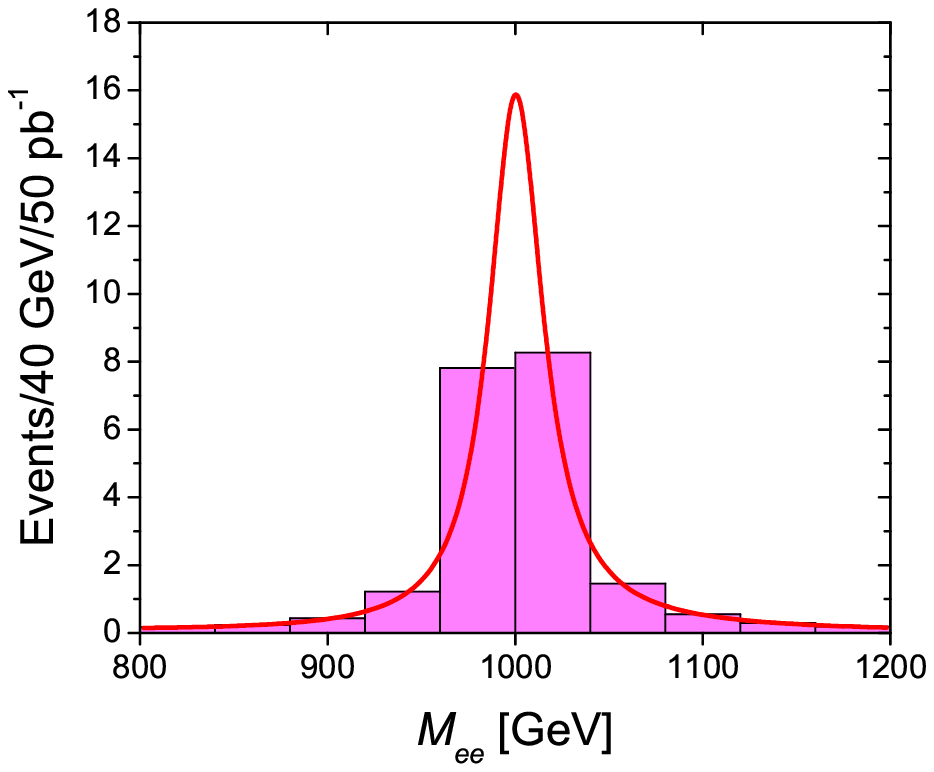,width=8cm}
\caption{\label{fig:resi} The invariant dilepton mass distributions
for the gauge $Z'$ boson (left) and the chiral excited $Z^*$ boson
(right).}
\end{figure}

The integrated luminosity around 50~pb$^{-1}$ can be reached this
year for two months LHC running at the initial luminosity of
$10^{31}$~cm$^{-2}$s$^{-1}$. Therefore, in the following
calculations we will use this value to estimate the  expected number
of events. The histograms on-peaks events with theoretical curves
are presented in Fig.~\ref{fig:resi}.
They correspond to $\sigma_{Z'}=0.45$~pb and $\sigma_{Z^*}=0.41$~pb,
and contain 22.5 and 20.6 events for the gauge and chiral bosons,
respectively.\footnote{The slight difference in the numbers is due
to different angular distributions and the applied cuts (see
\cite{0705.3944} for more details).} The irreducible Drell--Yan SM
background contributes only with $\sigma_{\rm bkgd}=5.75$~fb or 0.29
events. Therefore, the predicted numbers of the signal events
correspond to about 12$\,\sigma$ significant level of the
discovery.\footnote{The significances have been calculated according
the method presented in Appendix A of Ref.~\cite{CMS}, which follows
directly from the Poisson distribution.}

\begin{figure}[th]
\epsfig{file=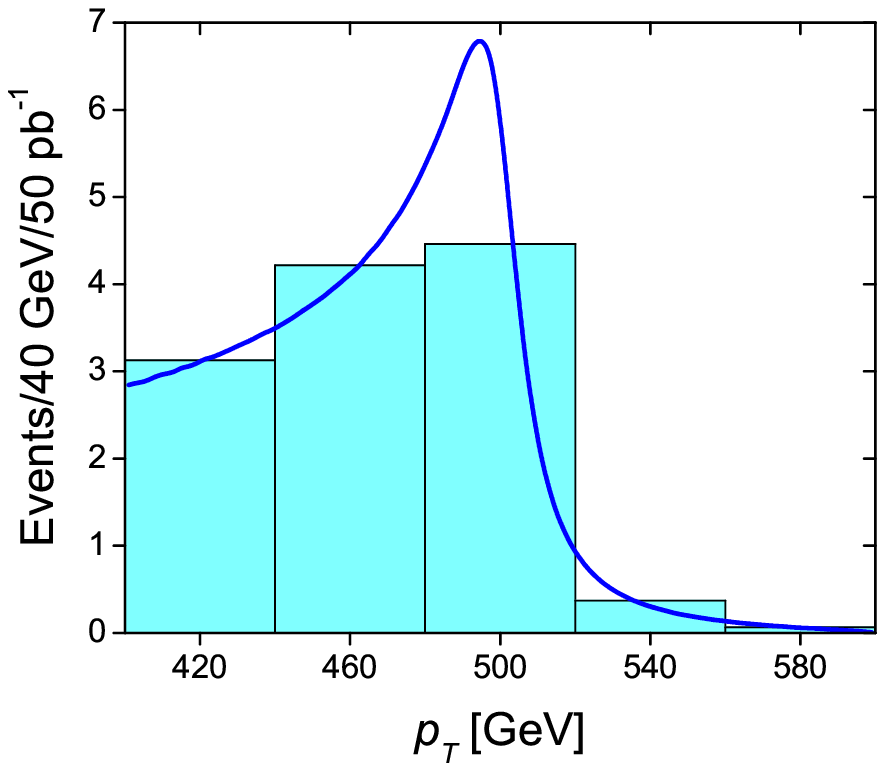,width=8cm}\epsfig{file=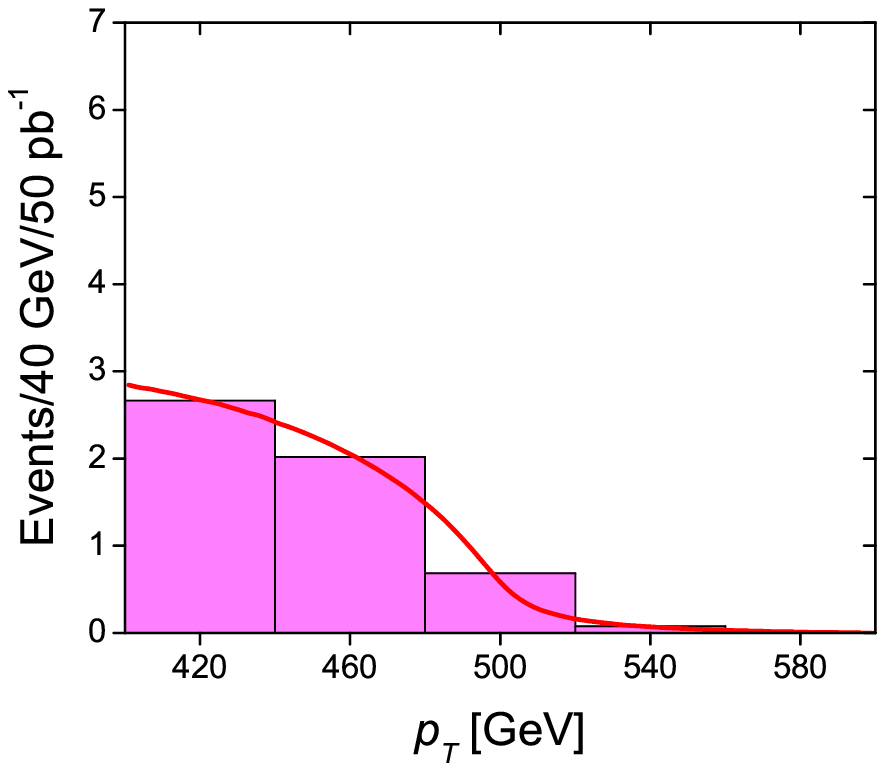,width=8cm}
\caption{\label{fig:rest} The differential distributions for the
gauge $Z'$ boson (left) and the chiral excited $Z^*$ boson (right)
as functions of the lepton transverse momentum.}
\end{figure}

In Fig.~\ref{fig:rest} we present the histograms and the theoretical
curves as functions of the lepton transverse momentum.
It can be seen that these distributions, in contrast to the previous
ones, are completely different one from another. Besides different
shapes they lead also to different ratios of the total number events
in the histograms in Fig.~\ref{fig:rest} to the total number events
under the corresponding peaks in Fig.~\ref{fig:resi}
\begin{equation}\label{rNtNtot}
    \frac{N_T^{Z'}}{N_{tot}^{Z'}}=0.54\pm 0.11,\hspace{2cm}
    \frac{N_T^{Z^*}}{N_{tot}^{Z^*}}=0.26\pm 0.10.
\end{equation}
Their difference is not statistically significant\footnote{The
variances have been calculated according to the binomial
distribution.} at this stage yet, but the last ratio shows a
tendency the transverse momentum distribution for the exited bosons
to be much softer than the analogous distribution for the gauge
bosons.

\begin{figure}[th]
\epsfig{file=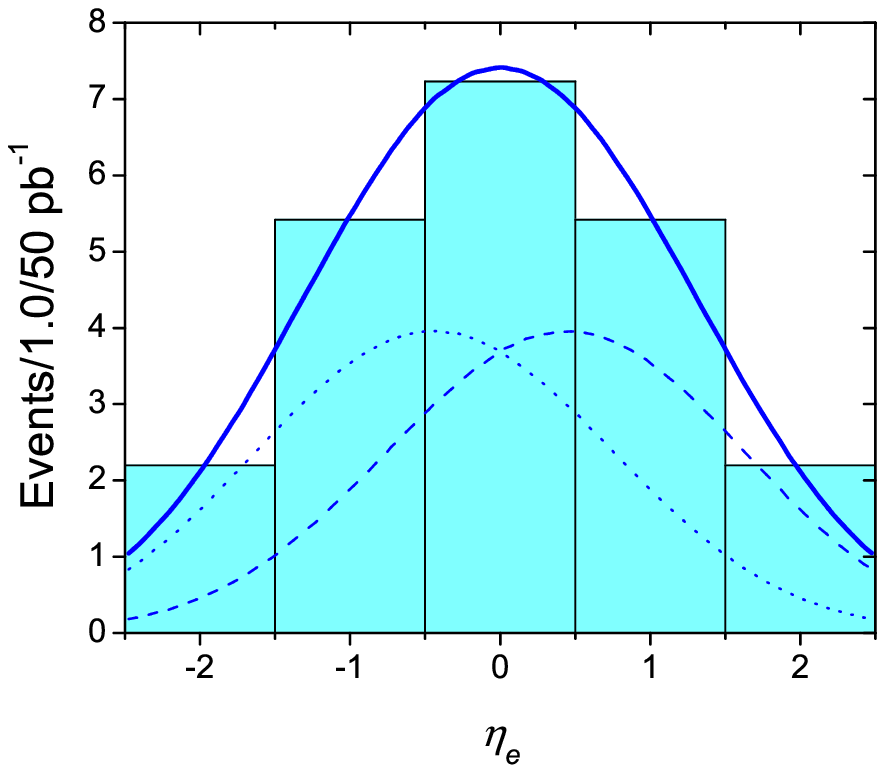,width=8cm}\epsfig{file=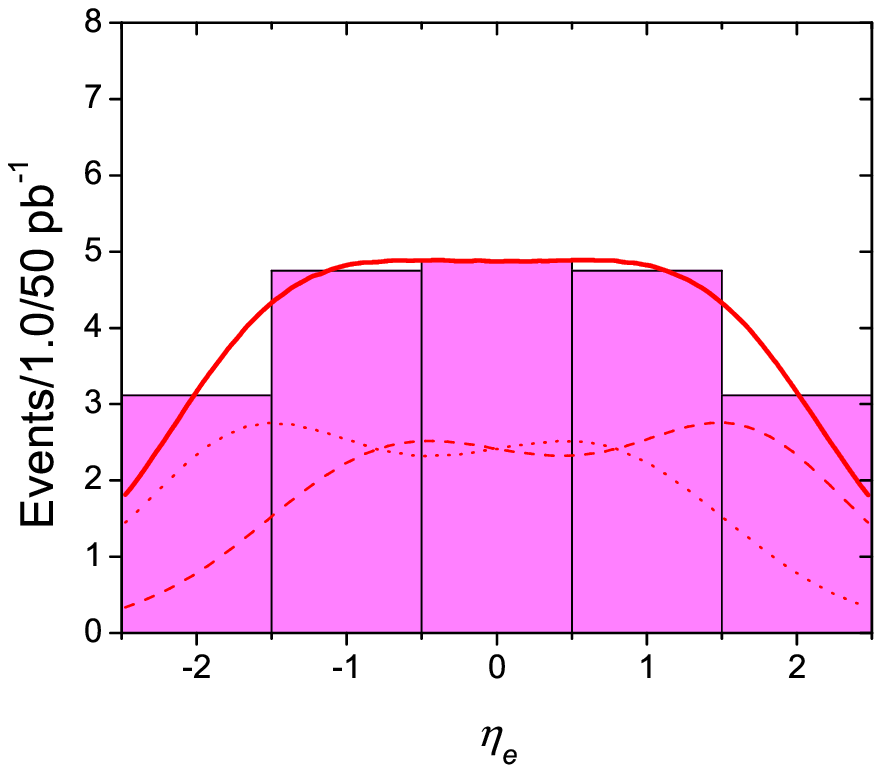,width=8cm}
\caption{\label{fig:reseta} The differential distributions for the
gauge $Z'$ boson (left) and the chiral excited $Z^*$ boson (right)
as functions of the lepton pseudorapidity. The corresponding
$d\bar{d}$ and $\bar{d}d$ contributions are shown as dashed and
dotted curves, respectively.}
\end{figure}

Let us turn now to the leptonic pseudorapidity distributions, which
are directly connected to the angular distributions (\ref{sV}) and
(\ref{sT}) up to the longitudinal boosts. To calculate the corresponding
distributions we need to fold
the parton cross sections with the parton luminosities $({\rm
d}N_{q\bar{q}}(x,\bar{x})/{\rm d}x\,{\rm d}\bar{x})$, where the
fractions of the quark and the antiquark momenta in the proton
\begin{equation}\label{xbarx}
    x=\frac{M_{\ell\ell}}{\sqrt{s}}\exp(+y);\hspace{1cm}
    \bar{x}=\frac{M_{\ell\ell}}{\sqrt{s}}\exp(-y)
\end{equation}
can be expressed through the center-of-mass energy, $\sqrt{s}$, the
invariant dilepton mass, $M_{\ell\ell}$, and the $\ell^-$ lepton
pseudorapidity, $y$, in the laboratory reference frame. Hence, the
parton luminosity is very sensitive to the latter. As far as the
neutral heavy bosons are produced at the LHC in $pp$ collisions, the
antiquarks come always from the sea and, in general, $\bar{x}<x$.
This means that the center of the $d\bar{d}$ distribution will be
shifted to positive $y$. The asymmetry is compensated by the
opposite shift of the $\bar{d}d$ distribution, which finally leads
to the symmetric form (see Fig.~\ref{fig:reseta}).\footnote{The
contributions of {\em up\/} type quarks are two orders of magnitude
smaller than the corresponding resonance contributions of {\em
down\/} type quarks in the on-peak region.}

According to Eq. (\ref{sT}), for the chiral bosons there exists a
characteristic plane, perpendicular to the beam axis in the parton
rest frame, where the emission of the final-state pairs is
forbidden. The nonzero probability in the perpendicular direction in
the laboratory frame is only due to the longitudinal boosts of the
colliding partons. This property is responsible for the additional
dips in the middle of the lepton pseudorapidity distributions for
the chiral bosons in contrast to the gauge ones (see the right panel
of Fig.~\ref{fig:reseta}). Therefore, the corresponding distribution
looks flatter at $\eta_\ell=0$ than the one for the gauge bosons.
However, this property is typical only for low values of the
center-of-mass energy, as $\sqrt{s}=10$~TeV or less, i.e.\ when the
corresponding parameter $\tau=M^2/s\ge 0.01$. At the nominal
$\sqrt{s}=14$~TeV the CERN LHC is sufficiently powerful to produce
heavy bosons of the mass M = 1 TeV with high longitudinal boosts,
which fill the dips, and the distributions for the gauge and chiral
bosons look almost similar (see \cite{CBB}).
\begin{figure}[th]
\epsfig{file=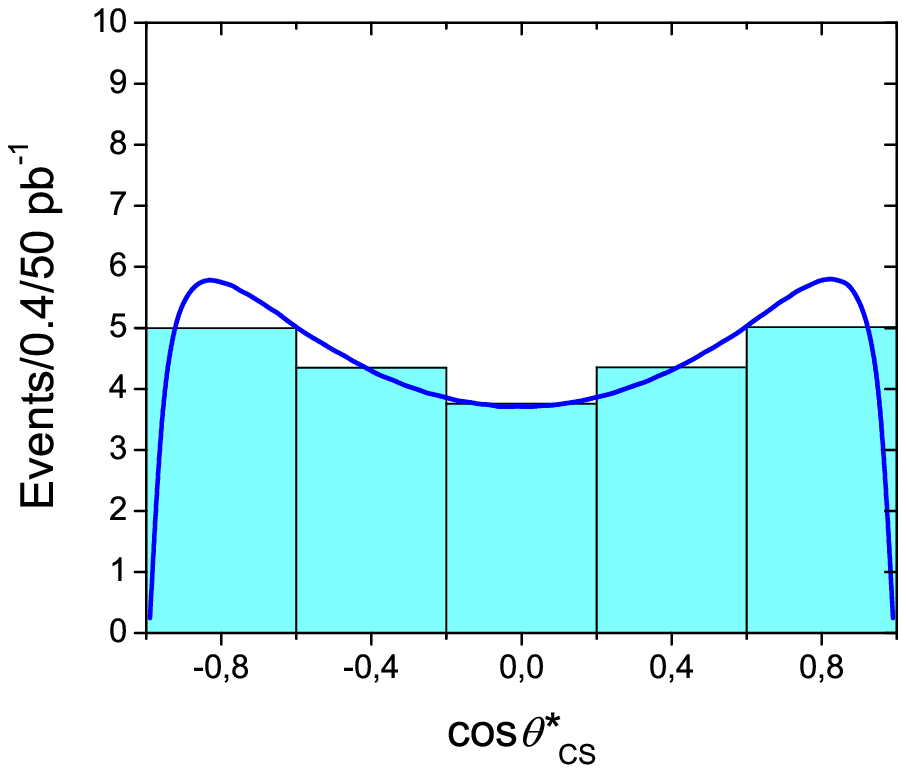,width=8cm}\epsfig{file=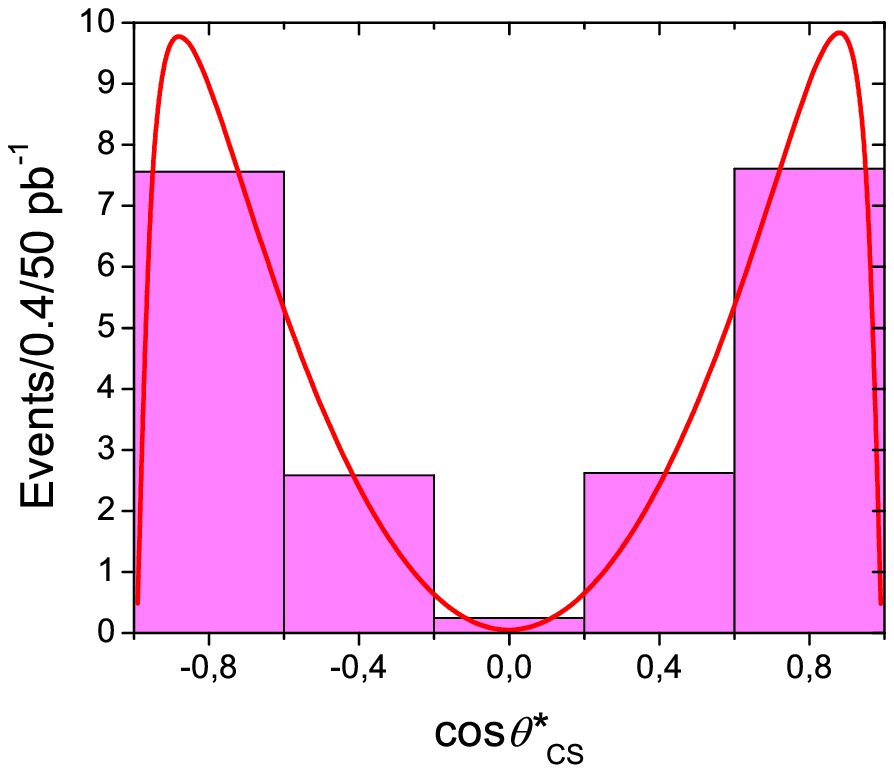,width=8cm}
\caption{\label{fig:rescos} The differential distributions of the
gauge $Z'$ boson (left) and the chiral excited $Z^*$ boson (right)
as functions of $\cos\theta^*_{\rm CS}$.}
\end{figure}

A crucial confirmation for the existence of the new interactions
(\ref{Z*ed}) should come from the analysis of the angular
distribution of the final leptons with respect to the boost
direction of the heavy boson in the rest frame of the latter (the
Collins--Soper frame). In Fig.~\ref{fig:rescos} we compare the
differential cross sections for the gauge $Z'$ boson and the chiral
excited $Z^*$ boson decaying to the lepton pairs with the invariant
mass 800~GeV~$<M_{\ell\ell}<$~1200~GeV as functions of
$\cos\theta^*_{\rm CS}$.

Instead of a smoother angular distribution for the gauge
interactions, a peculiar ``swallowtail'' shape of the chiral boson
distribution occurs with a profound dip at $\cos\theta^*_{\rm
CS}=0$. It is obvious that such form of the angular distribution
will lead to the large and negative value of the centre-edge
asymmetry $A_{CE}$, defined as~\cite{ACE}
\begin{equation}\label{ACE}
    \sigma A_{CE}=\int^{+\frac{1}{2}}_{-\frac{1}{2}}
    \!\frac{{\rm d}\sigma}{{\rm d}\cos\theta^*_{\rm CS}}\;{\rm d}\cos\theta^*_{\rm CS}-
    \left[\int^{+1}_{+\frac{1}{2}}
    \!\frac{{\rm d}\sigma}{{\rm d}\cos\theta^*_{\rm CS}}\;{\rm d}\cos\theta^*_{\rm CS}+
    \int^{-\frac{1}{2}}_{-1}
    \!\frac{{\rm d}\sigma}{{\rm d}\cos\theta^*_{\rm CS}}\;{\rm d}\cos\theta^*_{\rm CS}
    \right].
\end{equation}
Indeed, the corresponding values of the gauge and chiral bosons are
\begin{equation}\label{valuesACE}
    A^{Z'}_{CE}=-(0.11\pm 0.06),\hspace{2cm}
    A^{Z^*}_{CE}=-(0.69\pm 0.10),
\end{equation}
which differ at about 5$\sigma$ level. Therefore, the discovery of
the large negative $A_{CE}$ value will indicate the existence of the
heavy spin-1 boson with the new interactions (\ref{Z*ed}) already in
the first data. Neither scalars nor other particles possess such a
type of angular behavior.\footnote{The scalar and spin-2 particles
lead to the nonnegative $A_{CE}$ values and their signatures can be
rejected even at a  more significant level.}

\section{Conclusions}

In this paper we have considered the experimental signatures of the
chiral excited bosons, $Z^*$, and compared them to the gauge $Z'$
bosons. It has been stressed that the chiral bosons have a new
distinctive angular distribution, yet unknown by experimentalists.
It leads to an absence of the Jacobian peak in the transverse
momentum distribution and to a profound dip in the angular
distribution at the rest frame of the heavy chiral bosons. These
features could help to discriminate the chiral boson production from
a resonance production of other particles using the first LHC data.

\section*{Acknowledgements}
The author would like to thank I. Boyko for very useful
consultations. He is also grateful to V.G. Kadyshevsky, M.D. Mateev,
J.A. Budagov, N.A. Russakovich and V.A. Bednyakov for the support and
the fruitful cooperation.

This work was financially supported by the Brazilian agency CAPES.


\end{document}